\documentclass[%
reprint,
showpacs,preprintnumbers,showkeys,
 amsmath,amssymb,
 aps,
 prl,
superscriptaddress
]{revtex4-1}

\usepackage{graphicx}
\usepackage{dcolumn}
\usepackage{bm}
\usepackage{makecell}
\usepackage{multirow}
\usepackage{notes2bib}
\bibnotesetup{
note-name    = ,
use-sort-key = false
}

\begin{document}


\title{The infinite occupation number basis of bosons - solving a numerical challenge}

\author{Andreas Gei{\ss}ler}
\email{geissler@th.physik.uni-frankfurt.de}
\affiliation{Institut f\"ur Theoretische Physik, Goethe-Universit\"at, 60438 Frankfurt/Main, Germany}
\author{Walter Hofstetter}
\affiliation{Institut f\"ur Theoretische Physik, Goethe-Universit\"at, 60438 Frankfurt/Main, Germany}

\date{\today}

\begin{abstract}
In any bosonic lattice system, which is not dominated by local interactions and thus "frozen" in a Mott-type state, numerical methods have to cope with the infinite size of the corresponding Hilbert space even for finite lattice sizes. While it is common practice to restrict the local occupation number basis to $N_c$ lowest occupied states, the presence of a finite condensate fraction requires the complete number basis for an exact representation of the many-body ground state. In this work we present a novel truncation scheme to account for contributions from higher number states. By simply adding a single \textit{coherent-tail} state to this common truncation, we demonstrate increased numerical accuracy and the possible increase in numerical efficiency of this method for the Gutzwiller variational wave function and within dynamical mean-field theory.
\end{abstract}


\maketitle


\label{sec:Intro}

Applying any diagonalization-based method to bosonic lattice systems, which are not entirely in a Mott-type phase, often requires the use of a truncation scheme for the local Hilbert space. This is most evident for methods using variational wave functions, as for example the Gutzwiller state (GS) \cite{Gutzwiller1963,Gutzwiller1964,Gutzwiller1965,Rokhsar1991,Krauth1992} $\left| \psi \right\rangle = \prod_i \left| \psi_i \right\rangle$, because any numerical implementation requires a finite number of variational constants, which is realized by the choice of a truncation scheme. Related examples are the density matrix renormalization group (DMRG) \cite{White1992,Schollwock2005,Hallberg2006} and derived methods such as matrix product states (MPS) \cite{Ostlund1995,Verstraete2008,Schollwock2011,Orus2014}, projected entangled pair states (PEPS) \cite{Verstraete2008,Murg2007,Orus2014}, as well as time-evolving block decimation (TEBD) \cite{Vidal2003,Vidal2004,Verstraete2004,Zwolak2004}, which all require a truncation of the local occupation number basis to the $N_c$ lowest number states. The same is correspondingly true for bosonic single-impurity Anderson models (SIAM) \cite{Lee2010} as used in numerical renormalization-group (NRG) \cite{Glossop2007} approaches and dynamical mean-field theory (DMFT) \cite{Metzner1989,Georges1992,Georges1996,Kotliar2004,Anders2011a}. DMFT either relies on mapping a correlated many-body problem onto bosonic SIAMs \cite{Hubener2009a,Snoek2013} or directly solving the action via truncation-free stochastic methods \cite{Byczuk2008,Anders2011a}, such as the continous-time quantum Monte Carlo method \cite{Gull2011}. Nevertheless some effort has been made within DMRG, going beyond the simple truncation, by implementing an ``optimal phonon basis'' \cite{Zhang1998}, which is conceptually similar to our ansatz.

To a varying degree, all these methods will suffer from an insufficient truncation, while an increased basis size requires a corresponding increase in computing power. While matrix size can be limited independent of this truncation in DMRG methods, these usually describe states in terms of a locally truncated number basis. Therefore the cutoff $N_c$ also determines the possible overall truncation error. Furthermore, whenever solving a quantum impurity system by diagonalization, the corresponding matrices scale as $\prod_i M_i^2$, where $i$ represents internal degrees of freedom (DOF) and $M_i$ is the size of each corresponding Hilbert space, which require a truncation for bosonic DOF. The same relation is true for the variational GS, for which $i$ represents all sites and DOF under consideration.

As we will show for the cases of DMFT and GS, the use of a single additional variational basis state, which we denote as \textit{coherent-tail} state (CTS), can strongly increase the accuracy as compared to the common truncation scheme. Especially for DMFT the CTS is highly efficient: even strongly reduced Hilbert spaces suffice to well approximate the (quasi-)exact DMFT results, obtained by using a Hilbert space more than three times as large. Due to this reduction in computational complexity, this scheme is accompanied by a more than tenfold increase in numerical efficiency.

\label{sec:Sys}\textit{System} ---
In any numerical second quantized method, utilizing the grand canonical ensemble of an interacting Bose gas on a lattice, at some point it becomes necessary to approximate the infinite local Fock basis, to allow for results within a finite algorithm. As a test case, let us consider the basic Bose-Hubbard model \cite{Gersch1963,Fisher1989,Jaksch1998b}.
\begin{align}
H = -J \sum_{\left\langle i,j \right\rangle} (\hat{b}_i^{\dagger} \hat{b}_j + \hat{b}_j^{\dagger} \hat{b}_i) +  \frac{U}{2} \sum_i \hat{n}_i(\hat{n}_i-1) - \mu \sum_i \hat{n}_i \label{eq:BHmodel}
\end{align}
We use the common notation, where $\hat{b}_i$ ($\hat{b}^{\dagger}_i$) is the annihilator (creator) of a boson at site $i$, while $\hat{n}_i$ is the corresponding particle number operator $\hat{n}_i = \hat{b}^{\dagger}_i \hat{b}_i$. The parameters are the hopping amplitudes $J$ \cite{Bloch2008a}, the local Hubbard interaction $U$ \cite{Bloch2008a} -- tunable by Feshbach resonances \cite{Feshbach1958,Courteille1998,Inouye1998} -- and a chemical potential $\mu$, determining the total particle number.

Numerous techniques have been applied to investigate this model, ranging from the Gross-Pitaevskii equation (GPE) \cite{Polkovnikov2002,Kulkarni2015}, Bogoliubov theory \cite{Tikhonenkov2007,Kolovsky2007,Hugel2015} and variational mean-field methods such as GS \cite{Sheshadri1993,Buonsante2009} to more advanced techniques including Monte Carlo methods (MC) \cite{Capogrosso-Sansone2008,Kato2009,Pollet2013} and bosonic DMFT (BDMFT) \cite{Byczuk2008,Hubener2009a,Snoek2013,Anders2011a}. For numerical simulations in any of these methods, one needs to limit the infinite local Fock basis of bosons by a finite occupation number cutoff $N_c$. While $N_c$ can be arbitrarily high in principle, some methods require a comparatively low $N_c$, in order to limit the numerical effort. Let us now focus on BDMFT and GS, which become exact in both the atomic limit $J/U \rightarrow 0$ as well as the non-interacting limit $U/J \rightarrow 0$. In the last case the exact ground state can be written as a product of coherent states
$\left| \psi \right\rangle = \prod_i \left| \alpha_i \right\rangle$,
which also corresponds to the macroscopic condensate wave function $\Psi(i) = \left\langle \psi \left| \hat{b}_i \right| \psi \right\rangle$ solving the GPE. Despite some effort \cite{Krutitsky2011}, this correspondence is yet to be fully investigated.

For now we will focus on the intermediate superfluid regime, where for fixed chemical potential an increase in $J/U$ will result in an increasing mean particle number. In order to keep track of the ground state, one would generally need to include a proportionally increasing number of Fock states in any method that requires a $N_c$. This is true for both GS and BDMFT. In order to retain a small set of basis states, one should now switch to an optimized basis set, similar to \cite{Zhang1998}, but we also want to limit the computational cost. Therefore we propose a novel truncation scheme, where we replace only the highest included number state by the variational state $\left| \alpha_{N_c} \right\rangle$ as a linear combination of all remaining Fock states. Further requiring $\hat{b}\left| \alpha_{N_c} \right\rangle$ to be given as an exact linear combination of the new basis, thus reducing ``leakage'' out of the basis, yields the \textit{coherent-tail} state (CTS) $\left| \alpha_{N_c} \right\rangle$:
\begin{align}
\widetilde{\left| \alpha_{N_c} \right\rangle} = \sum_{n={N_c}}^{\infty} \frac{\alpha^{n}}{\sqrt{n!}} \left| n \right\rangle \label{eq:CTS}
\end{align}
This is a coherent state with the lower occupation numbers projected out. It therefore has to be normalized as $\left| \alpha_{N_c} \right\rangle = c_{N_c} \widetilde{\left| \alpha_{N_c} \right\rangle}$, with the factor $c_{N_c} = \left(\sum_{n={N_c}}^{\infty} \left|\alpha\right|^{2n}/n!\right)^{-1/2}$, to act as a proper basis state. This state extends the finite basis of ${N_c}$ Fock states $\left\lbrace 0,1,2,\ldots,{N_c}-1 \right\rbrace$, which in the following we denote as ${N_c}$-Fock basis, to $\left\lbrace 0,1,2,\ldots,{N_c}-1,\alpha_{{N_c}} \right\rbrace$. We would like to note that matrix elements within this basis will be as sparse as in the original representation, even in multi-component or cluster simulations \cite{Arrigoni2011,Luhmann2013}. We will now show how this soft bosonic truncation allows for significantly improved numerical accuracy in both GS and BDMFT and for a dramatically reduced calculation time at fixed accuracy within BDMFT.

\textit{Variational Gutzwiller state} --- 
We will first consider the GS in order to further introduce the method. GS uses the ansatz $\left| \psi_{\textrm{G}} \right\rangle = \prod_i \left| \psi_i \right\rangle$, where $\psi$ is usually written as a linear combination of the ${N_c}$-Fock basis states, while in our case this basis will be extended by the CTS. Due to the factorized wave function, the effective Hamiltonian has the following form

\begin{align}
H_{\textrm{G}} = -J \sum_{\left\langle i,j \right\rangle} (\hat{b}_i^{\dagger} \phi_j + \textrm{h.c.}) +  \frac{U}{2} \sum_i \hat{n}_i(\hat{n}_i-1) - \mu \sum_i \hat{n}_i \label{eq:GBHmodel}
\end{align}
where $\phi_i = \langle \hat{b}_i \rangle$. It is thus a set of local many-body problems coupled by the self-consistent fields $\phi_i$ (commonly called condensate order parameter). The ground state energy of this simplified Hamiltonian is found by variation of these fields. In a homogeneous system, where every site has $z$ nearest neighbours, and in the absence of spontaneous symmetry breaking, the problem reduces to a single variable $\phi$, thus further simplifying~\eqref{eq:GBHmodel}:

\begin{align}
H^{\textrm{local}}_{\textrm{G}} = -J z (\hat{b}^{\dagger} \phi + \phi^* \hat{b}) +  \frac{U}{2} \hat{n}(\hat{n}-1) - \mu \hat{n} \label{eq:singleGBHmodel}
\end{align}

\begin{figure}[h]
 \centering
  \includegraphics[width=0.49\columnwidth]{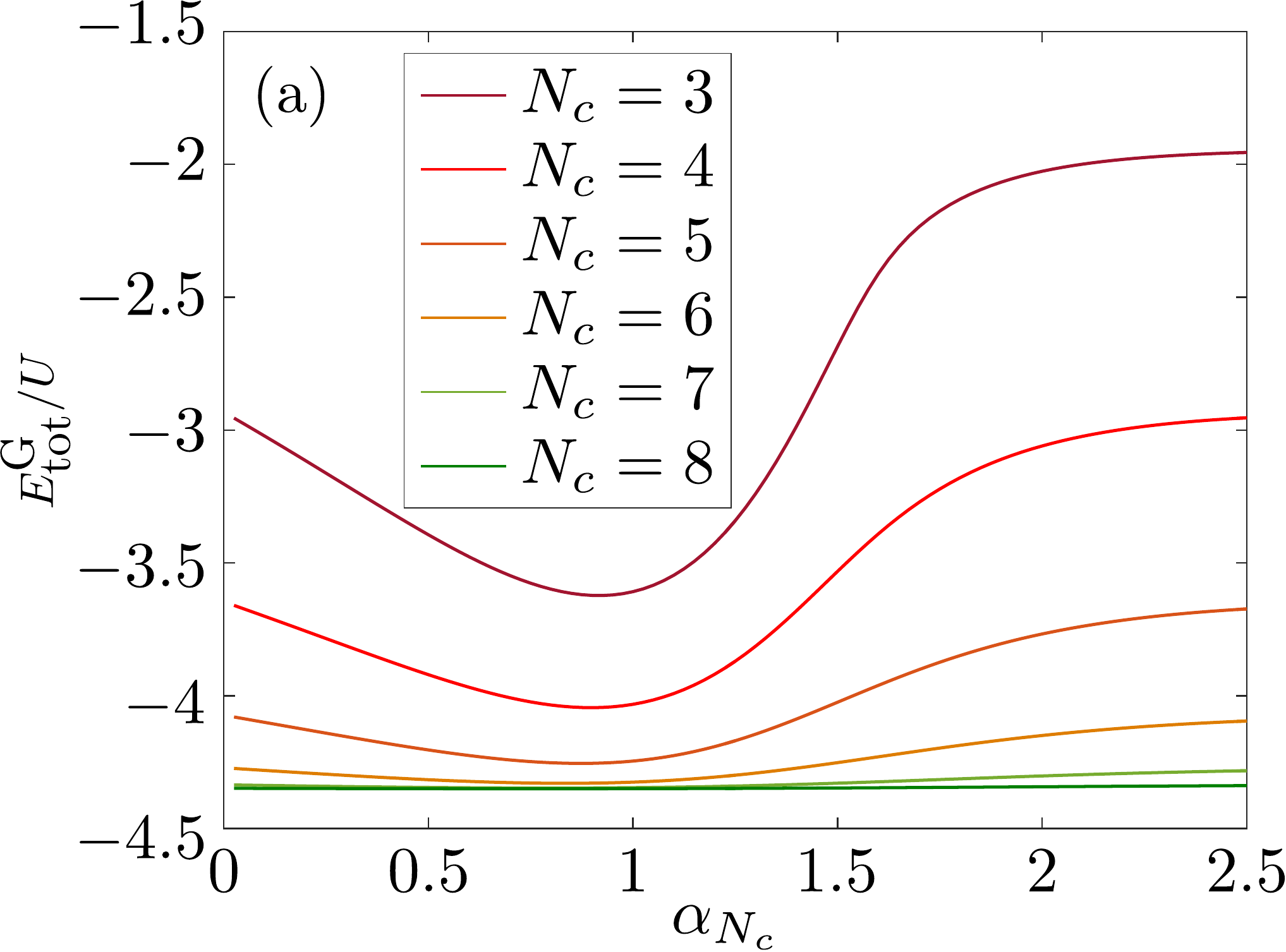}
  \includegraphics[width=0.48\columnwidth]{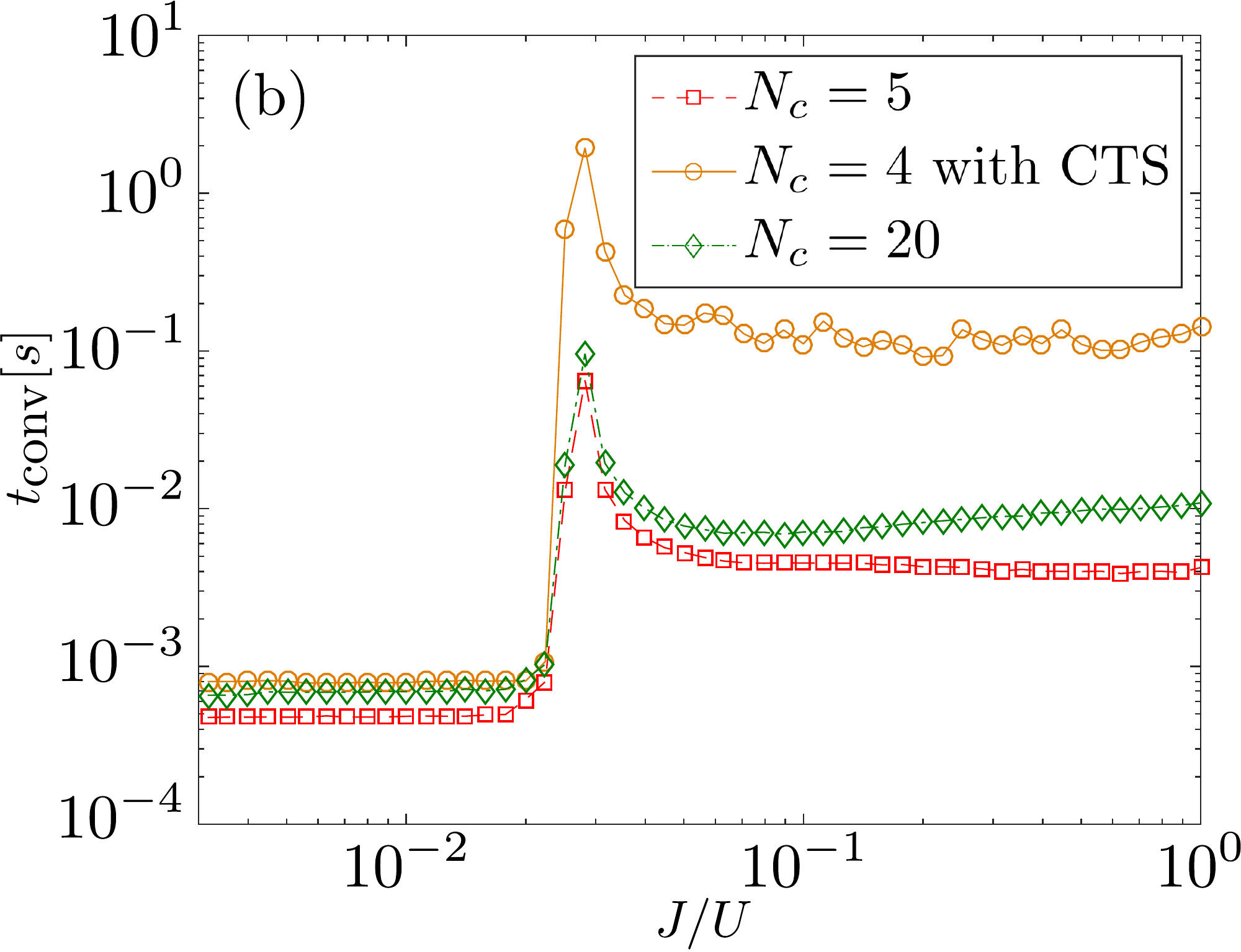}
  \includegraphics[width=0.49\columnwidth]{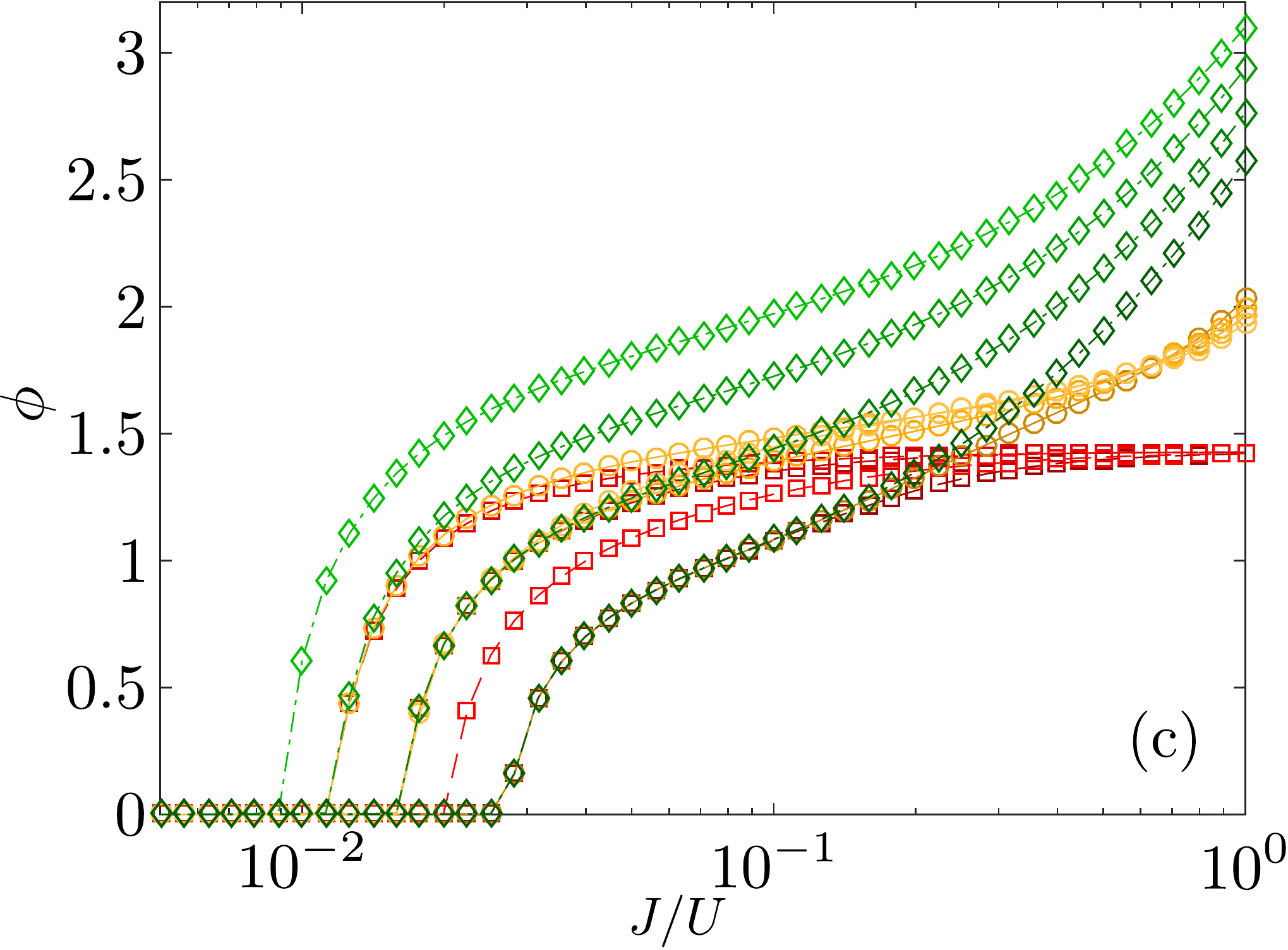}
  \includegraphics[width=0.46\columnwidth]{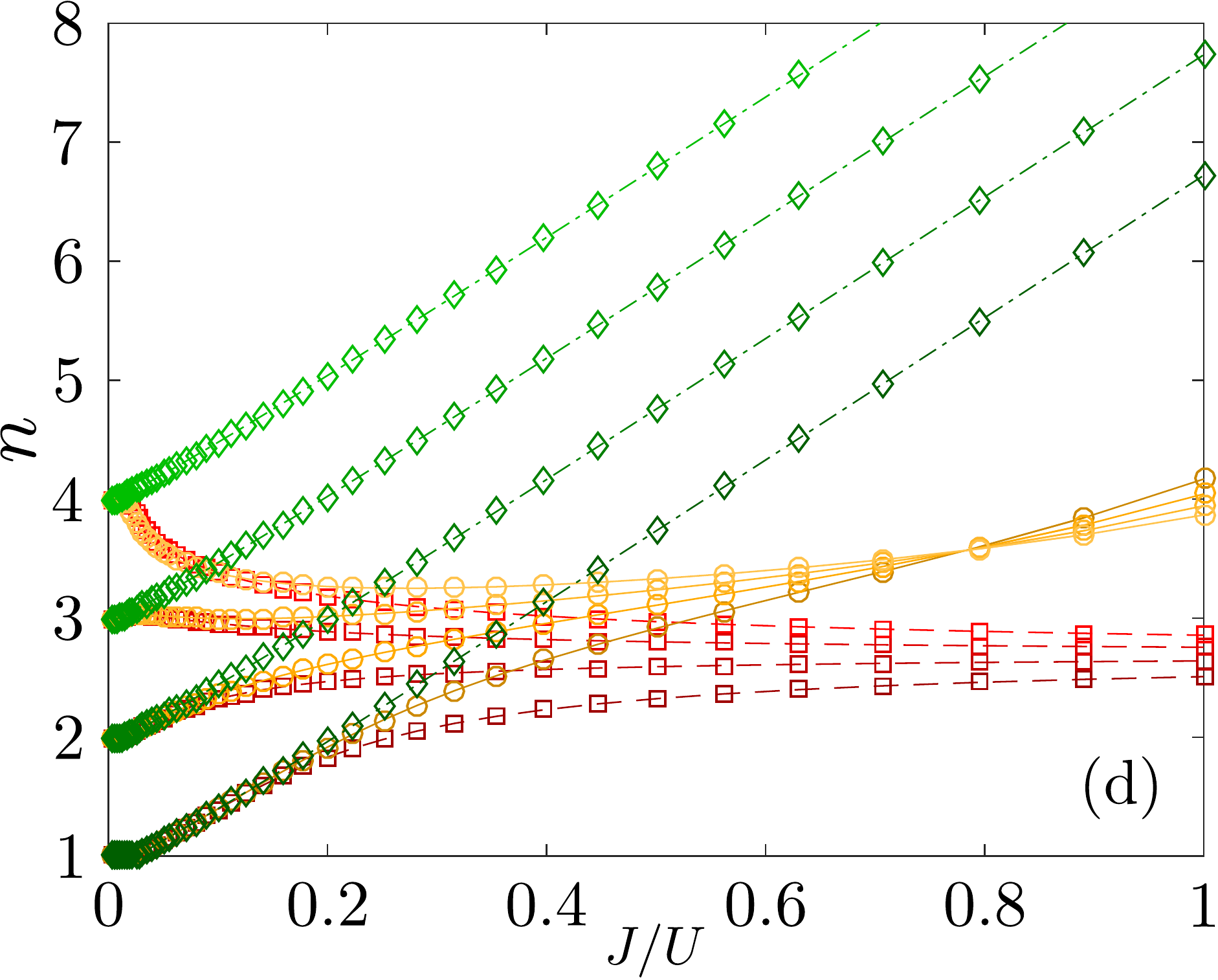}
 \caption{(a) Possible reduction of $E_{\textrm{tot}}^{\textrm{G}} = \left\langle \psi_{\textrm{G}} \left| H_{\textrm{G}}^{local} \right| \psi_{\textrm{G}} \right\rangle$ via variation of $\alpha_{N_c}$, for various truncations denoted by $N_c$, with $J/U = \mu/U = 0.4$ and $z=6$. (b) Convergence time of the GS for various truncation schemes, as given in the legend, and $\mu/U = 1.5$. Graphs (c,d) depict the expectation values of the observables $\langle \hat{b} \rangle$ and $\left\langle \hat{n} \right\rangle$ for various $\mu = 0.5,1.5,2.5,3.5$ (dark to bright colors; corresponding to the Mott lobes $n=1,2,3,4$) and truncations as in legend (b).}
 \label{fig:EGT}
\end{figure}

This problem can be solved in an arbitrary local basis, but any numerical implementation requires a truncation, for example to the common finite  ${N_c}$-Fock basis. In order to compare with numerical calculations in the CTS-extended basis, we furthermore need the following properties of the CTS
\begin{align}
\hat{b} \left| \alpha_{N_c} \right\rangle =& c_{N_c} \frac{\alpha^{N_c}}{\sqrt{({N_c}-1)!}} \left|{N_c}-1\right\rangle + \alpha \left| \alpha_{N_c} \right\rangle, \label{bCTS} \\
\hat{b}\hat{b} \left| \alpha_{N_c} \right\rangle =& c_{N_c} \frac{\alpha^{N_c}}{\sqrt{({N_c}-2)!}} \left|{N_c}-2\right\rangle \label{bbCTS} \\ &+ \alpha \left( c_{N_c} \frac{\alpha^{N_c}}{\sqrt{({N_c}-1)!}} \left|{N_c}-1\right\rangle + \alpha \left| \alpha_{N_c} \right\rangle \right) \nonumber
\end{align}
which are necessary to calculate all the additional matrix elements of $H^{\textrm{local}}_{\textrm{G}}$. Note that the CTS acts as the Fock state $\left| N_c \right\rangle$ for $\alpha \rightarrow 0$. Now one only needs to find the minimum of $E_{\textrm{tot}}^{\textrm{G}} = \left\langle \psi_{\textrm{G}} \left| H_{\textrm{G}}^{local} \right| \psi_{\textrm{G}} \right\rangle$, by simultaneous variation of both the physical parameter $\phi$ and the non-physical CTS-parameter $\alpha_{N_c}$. Since the final result has to be independent of the truncation scheme, a comparison for various $N_c$ and $\alpha_{N_c}$, at given values of $\mu/U = J/U = 0.4$, reveals the limited efficiency of the CTS (see Fig.~\ref{fig:EGT}). Thus we can now tell how a CTS-extended basis with reduced cutoff compares to a large ${N_c}$-Fock basis.

At any truncation level, if the CTS is added to the ${N_c}$-Fock basis, $E_{\textrm{tot}}^{\textrm{G}}$ is improved in comparison to a simple additional Fock state, corresponding to $\alpha_{N_c}\rightarrow 0$ in Fig.~\ref{fig:EGT}(a) (also Fig.~\ref{fig:EGT}(c,d)). One even improves upon the mean-field Mott transition, for Mott lobes $n = \langle \hat{n} \rangle = N_c$, at the limit of the Fock-basis with cutoff $N_c+1$ (see Fig.~\ref{fig:EGT}(c,d)). But due to the necessary optimization of $\alpha_{N_c}$, this comes at an additional computational cost (see Fig.~\ref{fig:EGT}(b)). The GS thus does not benefit much from the CTS, as far as computational effort is considered. On the other hand, as we will show, within BDMFT the CTS truncation scheme leads to a significant speed-up paired with the increased accuracy.

\textit{Bosonic dynamical mean-field theory} --- 
For BDMFT the CTS-extended Fock basis can be used in the (Anderson-)impurity solver within the self-consistency loop. Its implementation is  most straightforward in the exact diagonalization (ED) method. In that case the lattice Hamiltonian is mapped onto an effective local Hamiltonian \cite{Hubener2009a,Snoek2013}, which is an extended version of the GS Hamiltonian~\eqref{eq:GBHmodel}.

\begin{align}
 \mathcal{H}_{\mathrm{AIM}}^{\mathrm{eff}} =& \frac{U}{2} \hat{n}_0\left(\hat{n}_0-1\right) -\mu \hat{n_0} + \sum_l \epsilon_l \hat{a}_l^{\dagger} \hat{a}_l \label{eq:anderson} \\ 
 &-J\left( \hat{b}^{\dagger}_0 \left(\sum_{\langle i,0 \rangle} \langle \hat{b}_i\rangle_C\right)+ \hat{b}_0 \left(\sum_{\langle i,0 \rangle} \langle \hat{b}^{\dagger}_i\rangle_C\right) \right)\nonumber \\
 &+\sum_l \left( V_{l} \hat{a}^{\dagger}_l \hat{b}_0 + V_{l}^* \hat{a}_l \hat{b}^{\dagger}_0 +W_{l} \hat{a}_l \hat{b}_0+W_{l}^*\hat{a}^{\dagger}_l \hat{b}^{\dagger}_0\right)\nonumber
\end{align}
The additional terms including the annihilation (creation) operators $\hat{a}_l$ ($\hat{a}_l^{\dagger}$) describe effective bath orbitals which self-consistently mimic the action of the lattice sites surrounding the given site $j=0$ in the Hubbard model \eqref{eq:BHmodel}. They do so via the orbital energies $\epsilon_l$, normal hoppings $V_l$ and anomalous hoppings $W_l$. For an optimal representation of this action, increasing the number of bath orbitals is favorable over increasing bath truncations. They are therefore treated as hard-core bosons. The cavity expectation value $\left\langle . \right\rangle_C$ is computed in a system where the impurity site has been removed, which is required due to the mapping onto the effective model \cite{Hubener2009a,Snoek2013}. In the case of a homogeneous lattice gas, used here for benchmarking purposes, easily allowing for comparisons with truncations as high as $N_c = 20$, the term containing the self-consistent cavity order parameter simplifies to $\sum_i \left\langle b_i \right\rangle_C = z \cdot \phi_C$, where $z$ is the number of nearest neighbours, and $\phi_C$ is the cavity expectation value of the condensate order parameter.

\begin{figure}[h]
 \centering
  \includegraphics[width=0.95\columnwidth]{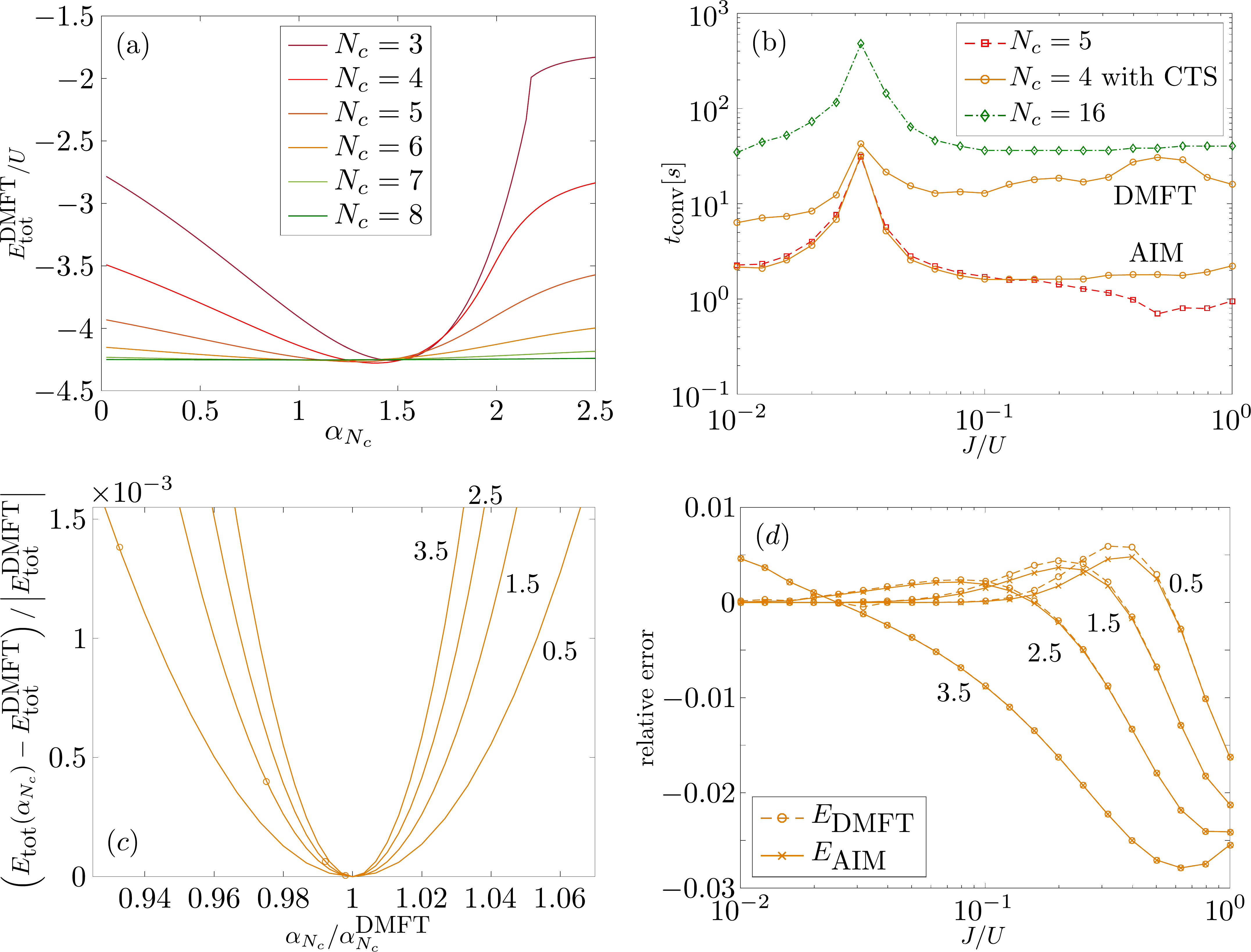}
 \caption{(a) Reduction of $E^{\textrm{DMFT}}_{\textrm{tot}}$ achieved by variation of the CTS via $\alpha_{N_c}$ for various values of $N_c$. Shown are self-consistent BDMFT results for $\mu / U = J/U = 0.4$ and $z=6$ (also used in (b,c,d)). (b) Comparison of convergence times of BDMFT for various truncation schemes, as given in the legend and annotation, with $\mu$ chosen as in (a). The two optimization schemes described in the text are compared in (c,d). (c) depicts the relative deviation of $E_{\textrm{tot}}(\alpha_{N_c})$ from its minimum, while the value found via minimization of $E_{\textrm{AIM}}$ is given by each marker. The corresponding simulations where performed for $J/U = 0.3162$. In (d) the resulting total energies, found by minimizing either $E_{\textrm{AIM}}$ or $E_{\textrm{DMFT}} = E^{\textrm{DMFT}}_{\textrm{tot}}$, are compared to the exact energies, as calculated for a regular cutoff $N_c = 20$. The used values of $\mu / U$ are given in each graph (c,d).}
 \label{fig:EAIM}
\end{figure}

\begin{figure}[h]
\includegraphics[width=0.99\columnwidth]{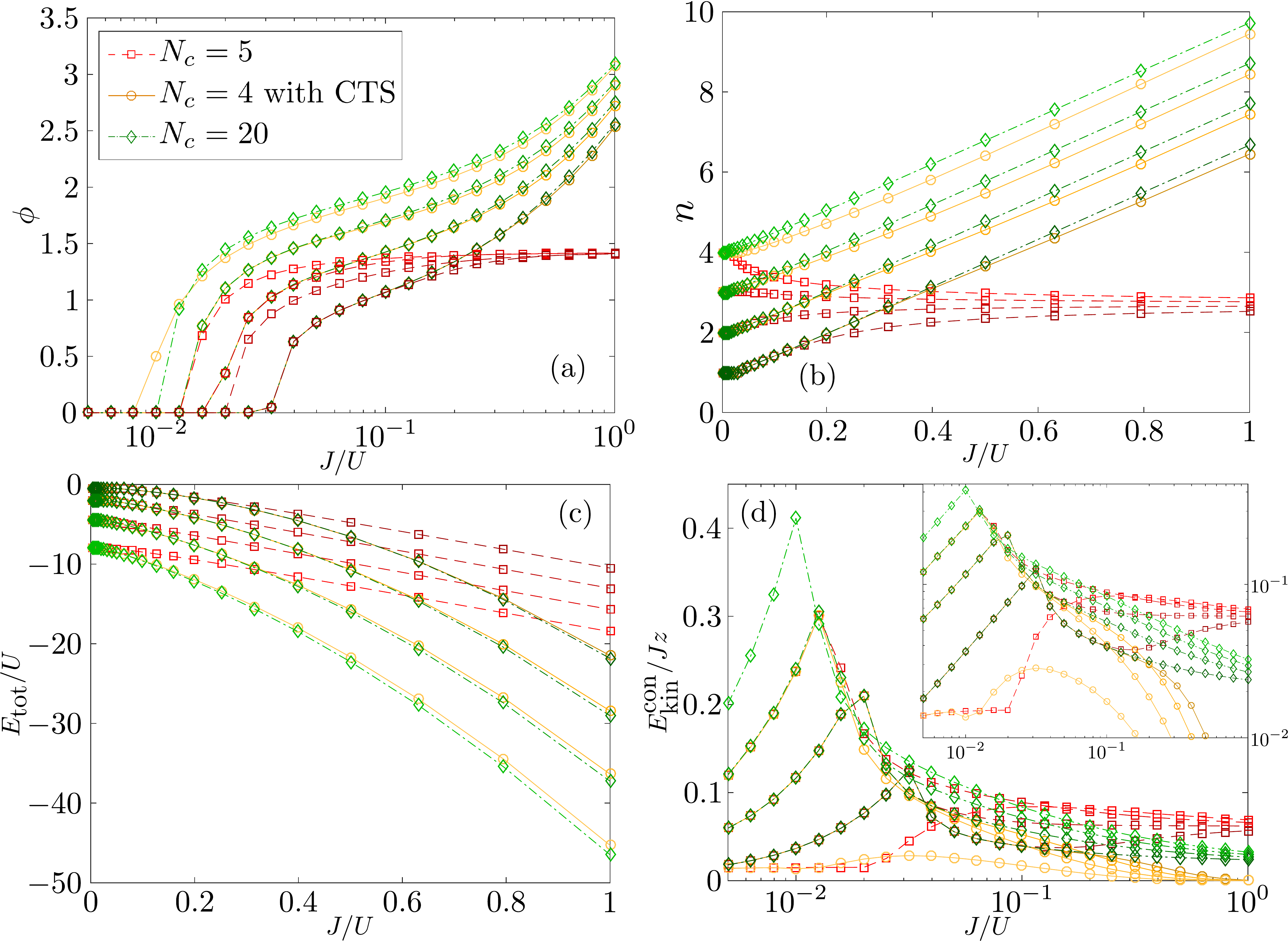}
 \caption{Results from converged BDMFT simulations, obtained for various truncation schemes, coded by colors and symbols shown in legend (a). Simulations were done for $\mu/U = 0.5,1.5,2.5,3.5$ (dark to bright colors) and $z=6$. Shown are the local observables $\phi$ (a), $n$ (b), the total energy per site $E_{\textrm{tot}} = \left\langle \hat{H} \right\rangle / L$ (c), where $L$ is the number of lattice sites, and the connected Green's function $G_c(t = 0)$ (d), as discussed in the main text \eqref{eq:Econkin}. The inset of (d) shows the same data plotted on a double log-scale for a better overview.}
 \label{fig:BDMFTobs}
\end{figure}

Within this implementation, a choice of $\alpha_{N_c}$ to compute the ground state of the full system, is efficiently obtained by minimizing the energy $E_{\textrm{AIM}} = {}_0\langle \mathcal{H}_{\mathrm{AIM}}^{\mathrm{eff}} \rangle_0$ for the lowest energy eigenstate of the self-consistent $\mathcal{H}_{\textrm{AIM}}^{\mathrm{eff}}$, with regard to the variational parameter $\alpha_{N_c}$. This yields the optimal representation for the low energy spectrum of $\mathcal{H}_{\textrm{AIM}}^{\mathrm{eff}}$, which determines the interacting Green's function in the Lehmann representation \cite{Snoek2013}. Another way of optimization would be the minimization of the self-consistently converged DMFT expectation value $E_{\textrm{tot}}(\alpha_{N_c}) = \langle H \rangle$ in relation to $\alpha_{N_c}$, where we define $E^{\textrm{DMFT}}_{\textrm{tot}}$ as this minimum. Let us emphasize, that the optimal $E_{\textrm{tot}}(\alpha_{N_c})$ should not depend on the choice of basis, so neither a variation in $N_c$ nor in $\alpha_{N_c}$ should result in a significant change of its self-consistent value, as indeed verified exemplary for $\mu/U = J/U = 0.4$ in Fig.~\ref{fig:EAIM}(a). Then the optimal CTS state allows for a remarkably good approximation of the total BDMFT energy, even at a very low Fock-space truncation $N_c$ (see Figs.~\ref{fig:EAIM}(a) and \ref{fig:BDMFTobs}(c)). 

A further look at the convergence times reveals the numerical benefit of replacing a large number of Fock-states (all those with $n \geq N_c$) by the single variational state $\left| \alpha_{N_c} \right\rangle$. We have simulated the Bose-Hubbard-model \eqref{eq:BHmodel} within BDMFT using a Bethe lattice with $z=6$ for $0 \leq \mu/U \leq 3.5$ and $0<J/U\leq1$. The convergence times for various truncation schemes are shown in Fig.~\ref{fig:EAIM}(b), for $\mu/U = 0.4$. Note the above $10$-fold decrease in convergence times when using the CTS-extended Fock basis, compared to the regular Fock basis with a high $N_c$, used for the (quasi-)exact solution. While this speed-up is only possible over the full range of parameters, when optimizing $E_{\textrm{AIM}}$, this simplified scheme leads to negligible deviations in the energy (as shown in Figs.~\ref{fig:EAIM}(c,d)). Also note the additional time loss of the CTS scheme, compared to a truncation of equal basis size at large $J/U$, which is due to the need to optimize $\alpha_{N_c}$, by finding either the minimum of $E_{\textrm{AIM}}$ or of $E_{\textrm{tot}}$, while the latter also requires multiple runs of fully converged DMFT simulations. So due to the minor differences, we now focus on results obtained via the first scheme.

Regarding physical observables, we have calculated local observables, such as the condensate order parameter $\phi = \langle \hat{b} \rangle$ and the occupation number $n = \left\langle \hat{n} \right\rangle$, as well as the non-local non-condensate fluctuations $G_c(t = 0) = - \left( \left\langle \hat{b}^{\dagger}_i \hat{b}_j \right\rangle - \left\langle \hat{b}^{\dagger}_i \right\rangle \left\langle \hat{b}_j \right\rangle \right)$, where $i$ and $j$ are nearest neighbours. This expression is more commonly denoted as the connected Green's function at equal times, which we can directly extract within BDMFT. Furthermore we also obtain the total energy $E_{\textrm{tot}}$ and the kinetic energy $E_{\textrm{kin}}^{\textrm{con}}$ due to the connected part of the Green's function, allowing for a comparison of the quality of different truncation schemes:
\begin{align}
\label{eq:Econkin}
E_{\textrm{kin}}^{\textrm{con}} = -Jz\left( \left\langle \hat{b}^{\dagger}_i \hat{b}_j \right\rangle - \phi^* \phi \right) = JzG_c(t = 0) 
\end{align} 

As is visible from the local observables as well as the total energy, replacing the highest Fock-state $\left| N_c \right\rangle$ by the CTS $\left| \alpha_{N_c} \right\rangle$ tremendously improves the results to almost the same accuracy as the (quasi-)exact result from the increased cutoff $N_c=20$ (see Fig.~\ref{fig:BDMFTobs}(a-c)). Remarkably, the CTS truncation even predicts the Mott transition for the Mott lobe $n=4$ (for $\mu/U=3.5$) almost exactly, as shown Fig.~\ref{fig:BDMFTobs}(a), while the regular cutoff $N_c=5$ fails to do so and both truncations also yield wrong $G_c(t = 0)$ (see Fig.~\ref{fig:BDMFTobs}(d)). The high accuracy of local observables is lost, just about where the occupation number $n$ exceeds $N_c$. Differences between the three cases can be seen most clearly in the non-condensed contribution to the kinetic energy \eqref{eq:Econkin}, due to non-local fluctuations described by the connected Green's function (see Fig.~\ref{fig:BDMFTobs}(d)). These have a monotonously decreasing tail for $J/U \rightarrow \infty$ in the exact solution. Obviously the ratio of these fluctuations to the condensate fluctuations of the BEC $\propto G_c(t=0) / \| \phi \|^2$ vanishes in this limit. But a hard and low truncation results in an artificially increased value of the non-local non-condensate fluctuations beyond certain values of $J/U$, while the CTS leads to the opposite behaviour, where the tail is damped more than in the exact result, thus suppressing non-condensate fluctuations early. This is likely a result of the CTS being more heavy-tailed than the Hubbard interaction would allow \cite{Krutitsky2011}. As non-condensate fluctuations only give a sub-leading contribution to the kinetic energy, it becomes clear why the CTS allows for the tremendous increase in accuracy and speed-up in numerical simulations compared to a simple high Fock space cutoff $N_c$, even for large $J/U$.

In conclusion, we have introduced a novel truncation scheme based on the CTS~\eqref{eq:CTS}, for which we demonstrated an increase in the numerical accuracy and computational efficiency of GS and BDMFT simulations. This increase was shown to be especially pronounced in BDMFT. Therefore the method allows for BDMFT simulations at much larger densities than before, but with reasonable computational effort. It is thus also a very promising method for accurate simulations of systems at higher filling per site.
Furthermore cluster-based methods \cite{Arrigoni2011,Luhmann2013} should especially benefit from this softened truncation, since the size of their Fock basis scales as $N_c^L$ with cluster size $L$. The concept of softening the hard cutoff, usually applied in the number basis, should thus more generally benefit a wide range of numerical simulations of bosonic lattice systems.

We would like to thank U. Bissbort and I. Vasi\'{c} for insightful discussions. Support by the Deutsche Forschungsgemeinschaft via DFG SPP 1929 GiRyd and the high-performance computing center LOEWE-CSC is gratefully acknowledged.

\bibliography{library}

\end{document}